\documentstyle[12pt,epsfig]{article}

\newcommand\as{\alpha_{\mathrm{S}}}

\newcommand\C{{\mathrm{c}}}
\newcommand\B{{\mathrm{b}}}
\newcommand\Z{{\mathrm{Z}}}
\newcommand\smfrac[2]{{\textstyle\frac{#1}{#2}}}

\def\cpc#1#2#3{Computer Phys.\ Comm.\ #1 (19#3) #2}

\def\np#1#2#3{Nucl.\ Phys.\ B#1 (19#3)~#2}
\def\pl#1#2#3{Phys.\ Lett.\ #1B (19#3) #2}

\def\zp#1#2#3{Zeit.\ Phys.\ C#1 (19#3) #2}

\def\be{\begin{equation}}
\def\ee{\end{equation}}
\def\ba{\begin{eqnarray}}
\def\ea{\end{eqnarray}}
\def\bann{\begin{eqnarray*}}
\def\eann{\end{eqnarray*}}
\def\benn{\begin{displaymath}}
\def\eenn{\end{displaymath}}
\def\nn{\nonumber}

\def\n{\, {\cal N}}
\def\tN{{\cal \widetilde{N}}}

\def\qqb{Q \bar Q}

\begin{document}

\begin{titlepage}
\begin{flushright}
  RAL-TR-98-042 \\ hep-ph/9805414
\end{flushright}
\par \vspace{10mm}
\begin{center}
{\Large \bf
Secondary Heavy Quark Pair Production \\[1ex]
in \boldmath$\mathrm{e^+e^-}$ Annihilation}
\end{center}
\par \vspace{2mm}
\begin{center}
{\bf D. J. Miller}\\
{and}\\
{\bf Michael H. Seymour}\\
\vspace{5mm}
{Rutherford Appleton Laboratory, Chilton,}\\
{Didcot, Oxfordshire.  OX11 0QX\@.  England.}
\end{center}
\par \vspace{2mm}
\begin{center} {\large \bf Abstract} \end{center}
\begin{quote}
\pretolerance 1000
  The multiplicity of heavy quarks from gluon splitting in
  $\mathrm{e^+e^-}$ annihilation has now been theoretically calculated
  and experimentally measured at LEP\@. However, the experimental
  measurement requires theoretical input for the shape of the
  multiplicity with respect to an event shape. In this paper we
  calculate the multiplicity of heavy quarks from gluon splitting in
  $\mathrm{e^+e^-}$ annihilation as a function of the heavy jet mass of
  the event, to next-to-leading logarithmic accuracy. We compare our
  result with Monte Carlo predictions.
\end{quote}
\vspace*{\fill}
\begin{flushleft}
  RAL-TR-98-042 \\ May 1998
\end{flushleft}
\end{titlepage}

\subsection*{Introduction}

Heavy quark production in $\mathrm{e^+e^-}$ annihilation can come from
two sources: from the hard interaction itself, $\mathrm{e^+e^- \to
  \qqb}$, and from the splitting of perturbatively produced gluons,
$\mathrm{e^+e^- \to q \bar q g \to q \bar q \qqb}$.  We call the latter
{\it secondary\/} heavy quarks. Their total rate is an infrared-safe
quantity, so can be calculated as an order-by-order perturbative
expansion in $\as$, starting at ${\cal O}(\as^2)$. The leading order
term in this expansion was calculated in Refs.~[\ref{Mike},\ref{Kuhn}].
At higher orders in $\as$, large logarithms arise, $\as^n
\log^{2n-1}(s/m_Q^2)$, potentially spoiling the convergence of the
perturbative series at high energies, $s \gg m_Q^2$. In Ref.~[\ref{early}] 
these (leading) logarithms were summed to all orders
in $\as$. The next-to-leading logarithmic terms, $\as^n
\log^{2n-2}(s/m_Q^2)$, were resummed in Ref.~[\ref{Mike}] yielding a result 
that is uniformly reliable for
all $s$. However, only the total multiplicity was calculated, retaining no
dependence on the jet kinematics. In this paper we perform a more
differential calculation, allowing the multiplicity to be calculated for
various event shapes.

Several experimental measurements of the secondary heavy quark
production rate have now been made. In Ref.~[\ref{OPAL1}], it was
extracted for charm quarks from a measurement of the $D^*$ fragmentation
function, and found to be more than a factor of two above the
expectation of Ref.~[\ref{Mike}], although with large systematic errors
coming from uncertainty in the fragmentation function of primary charm
quarks. References~[\ref{OPAL2}--\ref{ALEPH}] made less model dependent
measurements by selecting hard three-jet events, which enhances the
fraction of heavy quarks produced by the gluon splitting mechanism. In
general the measurements have been above the prediction of
Ref.~[\ref{Mike}], although within the range allowed by variations in
$\as$ and the quark mass.

For the parameters
\ba
\as&=&0.118,\\
m_\C&=&1.2 \mbox{~GeV},\\
m_\B&=&5.0 \mbox{~GeV},
\ea
we obtain results for the fractions of $\Z^0$ decays that
contain a secondary charm or bottom quark pair, of
\be
f_\C=2.007 \%,\quad
f_\B=0.175 \%.
\ee
Notice that these are somewhat higher than those of Ref.~[\ref{Mike}],
principally because we have used a different value for
$\as$\footnote{For our parameters, the calculation of Ref.~[\ref{Mike}]
  gives $f_\C=2.200 \%$ and $f_\B=0.207 \%$. The remaining difference is due
  to next-to-next-to-leading logarithms.}.  These should be compared
with the LEP values\footnote{For the bottom quark there has been only
  one measurement[\ref{DELPHI}], while for the charm quark, we have
  averaged the results of Refs.~[\ref{OPAL1},\ref{OPAL2}] and
  [\ref{ALEPH}], assuming that the systematic errors are uncorrelated.}
of
\be
  f_\C = (2.44 \pm 0.43)\%, \quad
  f_\B = (0.22 \pm 0.13)\%.
\ee
It should be noted that the summation of the large logarithms is
essential for this agreement. For example, the leading order result
alone for $f_\C$ is only $0.878$ for the same parameters.

However, these measurements are still somewhat model dependent, since
they rely on Monte Carlo event generators~[\ref{JETSET}--\ref{ARIADNE}]
to correct from the rate seen within the selected three-jet region, to
the total rate. Since in this paper we calculate this dependence
explicitly, it will become possible to directly compare experiment with
theory in the measured region, without the need to extrapolate to the
unmeasured region.

We begin by outlining the calculation of the resummed multiplicity,
retaining the dependence on the jet kinematics, and thereby allowing the
calculation of any event shape variable to be performed numerically. We
then present numerical results for a few quantities of interest.
Finally, we make some comparisons with the Monte Carlo event generators
that have been used previously, and draw some conclusions.

\subsection*{Calculation of the Resummed Multiplicity}

\subsubsection*{Fixed Order}
In order to separate the processes of primary and secondary heavy quark
production, it is necessary that the interference between them be zero,
or at least very small. Fortunately this is the case[\ref{Mike}].
For non-identical quarks coupled to a vector current, the interference
term will vanish by Furry's theorem if we assume that the charges of the
quarks are not measured (i.e.\ we do not distinguish quarks from
antiquarks). For an axial current, cancellations will occur between up-
and down-type quarks, leaving only the case where the ``light'' quark is
a bottom quark. This will only provide an effect of the order of $0.2\%$
of the secondary heavy quark rate[\ref{Kniehl}]. The contribution for
identical quarks is only slightly larger. In this case Furry's theorem
no longer applies but necessarily the quarks will be of the same
flavour~-- either bottom or charm, depending on the case in question,
and can be neglected.

Then, the leading order differential cross section for the production of
secondary heavy quarks, $\mathrm{\gamma^* \to q \bar q \qqb}$, is easily
calculated. In Ref.~[\ref{Mike}], this expression is integrated over the
heavy quark momenta, yielding the simple result given in Eq.~(2) of
Ref.~[\ref{Mike}]. In our case, since we must retain the jet kinematics,
we cannot do this integration since we would not correctly account for
events in which the heavy quarks fall into different hemispheres or
different jets. Instead we must retain the full unintegrated amplitude,
which does not have a compact form.
This expression is then integrated over the appropriate phase space by
Monte Carlo methods in which we can apply any event shape definition
desired.

We choose to calculate the resummed multiplicity of secondary heavy
quarks as a function of the heavy jet mass. Here we use the thrust
definition of heavy jet mass, i.e.\ we separate each event into two
hemispheres via the thrust axis and take the heavy jet mass to be the
larger of the two hemisphere masses. Our calculation is therefore easily
converted to give the differential distribution with respect to any {\em
  thrust-like\/} event shape.

\subsubsection*{Resummed Calculation}

There are two requirements to which we must conform in our calculation
of the logarithmic contribution. Firstly we must retain the exact
kinematics of the $q \bar qg$ production in order to be able to
accurately obtain the heavy jet mass for each event. We need not worry
about the exact kinematics of the heavy quarks, since the large
logarithms arise from the parts of phase space where they become
collinear and only the kinematics of the gluon need be considered
(although the exact kinematics of the heavy quarks are, of course,
included in the fixed order contribution). Secondly, we must include all
soft gluon emission from the light quarks and virtual gluon. These
emissions contribute to the heavy jet mass (by making the light quarks
massive) through large logarithms, which must be summed to all
orders using the coherent branching formalism.

Bearing these considerations in mind, we take the differential
multiplicity to be,
\ba
n_{e^+e^-}^{\qqb}(M_H^2,Q^2;Q_0^2) = \int dx_1 \, dx_2 \,
 dk_1^2 \, dk_2^2 \, dk_g^2 \, \frac{x_1^2+x_2^2}{k_{\perp}^2/Q^2}
f_q(k_1^2,k_{1 \, max}^2)f_q(k_2^2,k_{2 \, max}^2) \nn \\
n_g^{\qqb}(k_g^2,k_{\perp}^2;Q_0^2)
 \Theta(k_{\perp}^2-Q_0^2)\delta(M_H^2-h(x_1,x_2,k_1^2,k_2^2,k_g^2)).
\label{eq:goveq}
\ea
The above notation requires some clarification. As usual, $x_1$ and
$x_2$ are the energy fractions of the light quark and antiquark
respectively, and $k_1$, $k_2$ and $k_g$ are the four-momenta of the
quarks and gluon. The maximum value of $k_i^2$, $i=1,2 $, as constrained
by the phase space limits is $k_{i \, max}^2$. Also, $k_{\perp}^2$ is
the transverse momentum (squared) of the virtual gluon, given by,
\be
k_{\perp}^2 = (1-x_1+\epsilon_1-\epsilon_2)(1-x_2+\epsilon_2-\epsilon_1)Q^2,
\ee
where $Q^2$ is the centre-of-mass energy squared and
\be
\epsilon_i=\frac{m_i^2}{Q^2}, \quad i=1,2
\ee
are the rescaled (primary) quark masses (squared), which result from the
soft gluon emission. Here, $h(x_1,x_2,k_1^2,k_2^2,k_g^2)$ is the heavy
jet mass as a function of the exact kinematics of the quark, antiquark
and gluon.

Summation to all orders of the leading and next-to-leading logarithms is
included in the functions $f_q$ and $n_g^{\qqb}$. The function $f_q$ is
the quark jet mass distribution which has been calculated to
next-to-leading logarithmic accuracy in Ref.~[\ref{CTTW}]. More
explicitly, $f_q(k^2,Q^2) \, dk^2$ is the probability that a quark
created at a scale $Q^2$ gives rise to a jet with mass squared between
$k^2$ and $k^2+dk^2$. This function includes all soft gluon emission
from the light quarks and sums to all orders leading and next-to-leading
logarithms of $k_{i \, max}^2/k_i^2$, $i=1,2$.

The function $n_g^{\qqb}$ is the gluon jet mass distribution weighted by
the heavy quark pair multiplicity, and has not been calculated
elsewhere. In other words, $n_g^{\qqb}(k_g^2,k_{\perp}^2;Q_0^2) \,
dk_g^2$ is the number of heavy quark pairs within a gluon jet that was
formed at a scale $k_{\perp}^2$ and has a mass between $k_g^2$ and
$k_g^2+dk_g^2$. In calculating this quantity we must be sure to include
leading and next-to-leading logarithms of both $k_{\perp}^2/k_g^2$ and
$k_{\perp}^2/Q_0^2$.

Notice that the resolution scale at which the heavy quarks are resolved is,
\be
Q_0=2 \, m_Q^* = m_Q e^{5/6}.
\ee
Here, $m_Q^*$ is the heavy quark effective mass. It has been shown in
Ref.~[\ref{Mike}] that by using this effective mass we can neglect the
heavy quark mass in the decay of the gluon while maintaining the correct
leading and next-to-leading logarithms\footnote{This follows from the
  comparison of Eqs.~(16) and (17) of Ref.~[\ref{Mike}].}.

\noindent {\bf The Multiplicity Weighted Mass Distribution}

As in the case of the jet mass distribution, it is more convenient to
calculate the integrated distribution,
\be
N_g^{\qqb}(k^2,Q^2;Q_0^2)= \int_0^{k^2} dq^2 n_g^{\qqb}(q^2,Q^2;Q_0^2).
\ee
Physically this is the number of $\qqb$ pairs resolved in gluon jets of
mass squared {\em less than}~$k^2$. It can be derived from
$N_g^g(k^2,Q^2;Q_0^2)$, the multiplicity of gluons within gluon jets of
mass squared less than $k^2$, by integrating over the kernel for the
splitting $\mathrm{g \to \qqb}$, i.e.\ $P_{qg}$. Therefore, we have,
\ba
N_g^{\qqb}(k^2,Q^2;Q_0^2) &=& \int_{Q_0^2}^{k^2} \frac{dq^2}{q^2} \int_0^1 dz
\frac{\as(q)}{2\pi} P_{qg}(z) N_g^g(k^2,Q^2;q^2) \nn \\
&=& \frac{1}{3} \int_{Q_0^2}^{k^2} \frac{dq^2}{q^2}
\frac{\as(q)}{2\pi} N_g^g(k^2,Q^2;q^2).
\label{eq:intN}
\ea
$N_g^g$ has been derived in Ref.~[\ref{us}], and is given by,
\be
N_g^g(k^2,Q^2;Q_0^2)=F_g(k^2,Q^2) \left\{ \n_g^g(k^2;Q_0^2)
+ C_A \left(I^g(k^2;Q_0^2)-I^g(k^4/Q^2;Q_0^2) \right) \right\}
\ee
where,
\ba
I^g(k^2;Q_0^2)&=&\Theta(z_k-z_0)\frac{1}{C_A}
 \left[ \n^+(z_0,z_k)-1+2B \tN(z_0,z_k) \right. \nn \\
&& \left. \hspace{3cm} - \smfrac12C
 \left(2 \, \tN(z_0,z_k)-z_k^2/z_0^2+1 \right) \right],
\ea
and $\n_g^g$ is the usual multiplicity of gluons within a gluon[\ref{CDFW}],
\be
\n_g^g(k^2;Q_0^2) = \n^+(z_0,z_k) - C \tN(z_0,z_k).
\ee
Here, we have used the variables introduced by Catani et
al.[\ref{CDFW}], where,
\be
z_0^2=\frac{32 \pi C_A}{b^2 \as(Q_0)}, \quad \quad \quad
z_k^2=\frac{32 \pi C_A}{b^2 \as(k)}.
\ee
The functions, $\n^+$ and $\tN$, are defined in terms of Bessel functions by,
\ba
\n^+(z_0,z_k)&=& z_k \left( \frac{z_0}{z_k} \right)^B
 \left[ I_{B+1}(z_k)K_{B}(z_0)+K_{B+1}(z_k)I_{B}(z_0) \right], \nn \\
\tN(z_0,z_k)&=& \left( \frac{z_0}{z_k} \right)^B
 \left[ I_{B}(z_k)K_{B}(z_0)-K_{B}(z_k)I_{B}(z_0) \right],
\ea
and the parameters $B$ and $C$ are given by,
\be
B=\frac{1}{b} \left( \frac{11}{3} C_A
 +\frac{2N_f}{3}-\frac{4C_FN_F}{3C_A} \right), \quad \quad \quad
 C=\frac{8}{3} \frac{N_f}{b} \frac{C_F}{C_A},
\ee
where $b$ is the first coefficient of the $\beta$-function,
\be b=\frac{11}{3}C_A-\frac{2}{3}N_f. \ee
The gluon jet mass fraction, $F_g(k^2,Q^2)$, is the probability that a
jet formed at scale $Q^2$ will have a mass of less than $k^2$. This is
the integrated version of $f_g(k^2,Q^2)$ and is given, in
Ref.~[\ref{CTTW}], by,
\ba
F_g(k^2,Q^2) &=& \int_0^{k^2} dq^2 f_g(q^2,Q^2) \nn \\
&=&\exp \left\{ C_A \log \left( \frac{Q^2}{k^2} \right) f_1
\left( \frac{\as(Q)}{4\pi} b \log \left( \frac{Q^2}{k^2} \right) \right)
+ \frac{b}{2} f_2
\left(  \frac{\as(Q)}{4\pi} b \log \left( \frac{Q^2}{k^2}
 \right) \right) \right\}, \nn \\
\ea
where,
\ba
f_1( \lambda ) &=& \frac{2}{b \lambda}
\left[(1-2 \lambda ) \log \left( \frac{1}{1-2 \lambda}
 \right)-2(1- \lambda )\log \left( \frac{1}{1- \lambda} \right)
\right], \\
f_2( \lambda ) &=& \frac{2}{b} \log \left( \frac{1}{1- \lambda} \right).
\ea
Notice that the result of Ref.~[\ref{CTTW}] includes leading and
next-to-leading logarithms within $\log F_g$, whereas the expression
above is accurate only to leading and next-to-leading logarithms within
$F_g$ itself.

The integrals required for the evaluation Eq.~(\ref{eq:intN}) have been
derived in Ref.~[\ref{Mike}] and are, to next-to-leading logarithmic
accuracy,
\ba
\int_{z_0}^{z_k} \frac{dz}{z} \n^+(z,z_k) &=& \tN_1(z_0,z_k) - \frac{2}{z_0^2}
 \left( 1-B \left(1-\frac{z_0}{z_k} \right) \right)
 \n^+(z_0,z_k) + \frac{2}{z_k^2}, \\
\int_{z_0}^{z_k} \frac{dz}{z} \tN(z,z_k)&=&
 \frac{1}{z_0z_1}\n^+(z_0,z_k) - \frac{1}{z_k^2}.
\ea
This gives,
\be
N_g^{\qqb}(k^2,Q^2;Q_0^2)=F_g(k^2,Q^2) \left\{ \n_g^{\qqb}(k^2;Q_0^2)
+ C_A \left(I^{\qqb}(k^2;Q_0^2)-I^{\qqb}(k^4/Q^2;Q_0^2) \right) \right\},
\ee
with,
\ba
I^{\qqb}(k^2;Q_0^2) &=& \Theta(z_k-z_0) \frac{4}{3b \, C_A}
\left\{ \tN_1(z_0,z_k) + \left( 2(B-1)\frac{1}{z_0^2}- C\frac{1}{z_0z_k}
 \right) \n^+(z_0,z_k) \right. \nn \\
&& \left. - \frac{1}{z_k^2} \left( 2B-2-C \right) - \smfrac14 (C+2)
 \log \left( \frac{z_k^2}{z_0^2} \right) - \frac{C}{4}
 \left( 1- \frac{z_k^2}{z_0^2} \right) \right\},
\ea
and,
\be
\n_g^{\qqb}(k^2;Q_0^2) = \frac{4}{3b} \left\{ \tN_1(z_0,z_k) +
 \left( \frac{2}{z_0^2} (B-1) - \frac{1}{z_0z_k}(2B+C) \right)
 \n^+(z_0,z_k) + \frac{1}{z_k^2}(C+2) \right\}.
\ee
In the above and following $\n_1$ is defined as $\n$ but with $B$
replaced by $B-1$.

Finally $n_g^{\qqb}$ is obtained by differentiation with respect to the
jet mass, yielding the result,
\[
n_g^{\qqb}(k^2,Q^2;Q_0^2)= \frac{1}{k^2} F_g(k^2,Q^2)
 \left\{ \n_g^{\prime \qqb}(k^2;Q_0^2)
+C_A \left(I^{\prime \qqb}(k^2;Q_0^2)- 2 \, I^{\prime \qqb}(k^4/Q^2;Q_0^2)
 \right)
\right\} \] \nopagebreak \vspace{-0.3cm} \be
+ \frac{d}{dk^2} \left( \log F_g(k^2,Q^2) \right) N_g^{\qqb}(k^2,Q^2;Q_0^2),
\ee
\pagebreak
with,
\ba
\n_g^{\prime \qqb}(k^2;Q_0^2)&=&
\frac{16C_A}{3b^2} \left\{ \frac{1}{z_k^2} \n^+_1(z_0,z_k)
- \left( \frac{2}{z_0^2}(B-1) - \frac{1}{z_0z_1}(2B+C) \right)
\tN(z_0,z_k) \right\}, \nn \\
I^{\prime \qqb}(k^2;Q_0^2)&=&
 \frac{16}{3b^2} \left\{ \frac{1}{z_k^2} \left( \n^+_1(z_0,z_k)-1 \right)
+ \left( \frac{2}{z_0^2}(B-1)-C\frac{1}{z_0z_k} \right)
 \tN(z_0,z_k) \right. \nn \\
&& \quad \quad \left. - \frac{C}{2} \frac{1}{z_k^2}
 \left( 1- z_k^2/z_0^2 \right) \right\},
\ea
and,
\be
k^2 \frac{d}{dk^2} \left( \log F_g(k^2,Q^2) \right)=
\frac{4C_A}{b} \log \left( \frac{1-\lambda}{1-2\lambda}
\right)
-\left( \frac{\as}{2\pi} \right) \frac{b}{2} \frac{1}{1-\lambda},
\ee
where,
\be \lambda=\frac{\as(Q)}{2\pi} b \log(Q/k). \ee

Notice that the matching to tree level is particularly simple. We only
require the expansion of $n_g^{\qqb}$ to order $\as$, which is,
\be
 n_g^{\qqb}(k^2,Q^2;Q_0^2)= \frac{1}{3k^2} \frac{\as}{2\pi} + {\cal O}(\as^2).
\ee

\vspace{-5pt}
\subsubsection*{Calculation of the Background}
\vspace{-5pt}
The background to secondary heavy quark production in $\mathrm{e^+ e^-}$
annihilation, i.e.\ primary heavy quark production, can be estimated by
standard three-jet production, since the mass effects will be small. For
the analysis of the heavy jet mass, the fixed order contribution to
this background is given by,
\ba
\frac{1}{\sigma_0}\frac{d\sigma^{(1)}(\tau)}{dM_H^2} &=& \frac{1}{M_H^2} C_F
\frac{\as(Q)}{2\pi} \left\{ -4 \log \tau -3 +6 \tau \log \tau - 4
\frac{\tau^2}{1-\tau} \log \left(1- \tau \right) \right.\nn \\
&&\left. + \frac{8}{1-\tau} \log \left(1- \tau \right) - 6 \tau
\log \left(1- 2 \tau \right) -\frac{4}{1-\tau} \log \left(1-2 \tau \right)
+9 \tau^2 \right\}
\ea
where $\sigma^{(1)}$ is the ${\cal O}(\as)$ contribution to the cross
section, $\sigma_0$ is the Born cross section, $\tau = M_H^2/Q^2$ and
$M_H$ is the heavy jet mass.

Of course, the large leading and next-to-leading logarithms must also be
included and are given by,
\be
\frac{1}{\sigma_0}\frac{d\sigma^{(\mathrm{logs})}(\tau)}{dM_H^2} =
2f_q(M_H^2,Q^2)F_q(M_H^2,Q^2).
\ee
Expanding this to ${\cal O}(\as)$ gives the first two terms of the fixed
order piece,
\be
\frac{1}{\sigma_0}\frac{d\sigma^{(\mathrm{logs})}(\tau)}{dM_H^2} =
\frac{1}{M_H^2} C_F \frac{\as(Q)}{2\pi} \left\{ -4 \log \tau -3 \right\}
+ {\cal O}(\as^2).
\ee
Therefore it is clear that matching with the fixed order will result in
the full answer,
\ba
\frac{1}{\sigma_0}\frac{d\sigma(\tau)}{dM_H^2} &=&
2f_q(M_H^2,Q^2)F_q(M_H^2,Q^2) + \frac{1}{M_H^2} C_F
\frac{\as(Q)}{2\pi} \left\{6 \tau \log \tau - 4
\frac{\tau^2}{1-\tau} \log \left(1- \tau \right) \right. \nn \\
&&\left. + \frac{8}{1-\tau} \log \left(1- \tau \right) - 6 \tau
\log \left(1- 2 \tau \right) -\frac{4}{1-\tau} \log \left(1-2 \tau \right)
+9 \tau^2 \right\}.
\ea
This can be seen plotted in Fig.~\ref{fig:heavy}.
\pagebreak

\subsection*{Numerical Results}

For all the distributions we show, we concentrate on their shape,
normalised to the number of secondary heavy quarks, rather than on the
total rate. We use the $\as$ and quark mass values quoted earlier.

\begin{figure}[thbp]
\begin{center}
~\epsfig{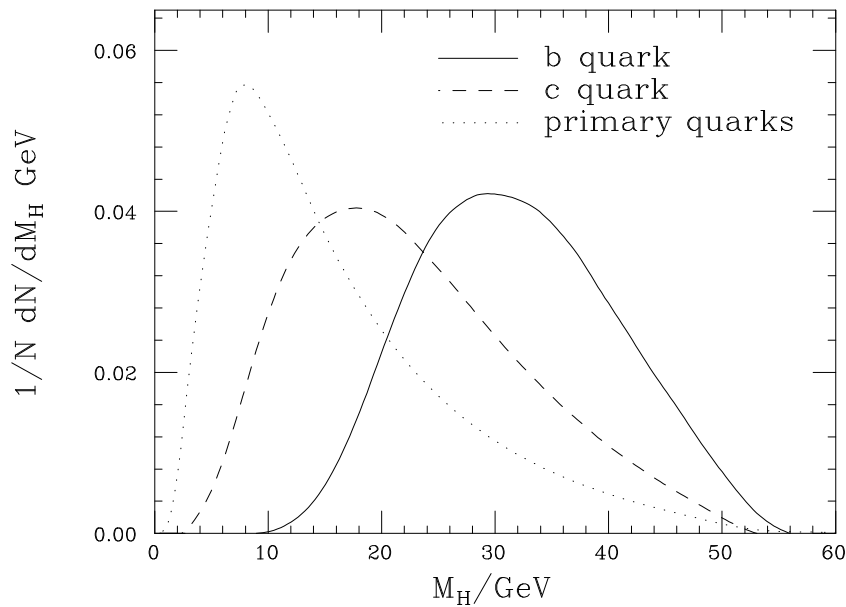}
\end{center}
\caption[heavy]{The multiplicity of heavy quark pairs in $\Z^0$ decays
  as a function of the heavy jet mass, normalised to the number of
  secondary heavy quarks. The shapes for b quark (solid curve), and c
  quark (dashed) pairs are compared to the three-jet background
  (dotted). It is clear that secondary heavy quark production can be
  distinguished from the background by the shape of the heavy jet mass
  distribution.}
\label{fig:heavy}
\end{figure}

We present the heavy jet mass distribution for $\surd s=m_\Z$ in
Fig.~\ref{fig:heavy}.  This is closely related to the jet mass
difference, $M_H-M_L$, which was the event shape used to fit $f_c$ in
Ref.~[\ref{ALEPH}]. We see that the heavy jet mass provides a good
discriminator of events with secondary heavy quarks from the three-jet
background.

\begin{figure}[thbp]
\begin{center}
~\epsfig{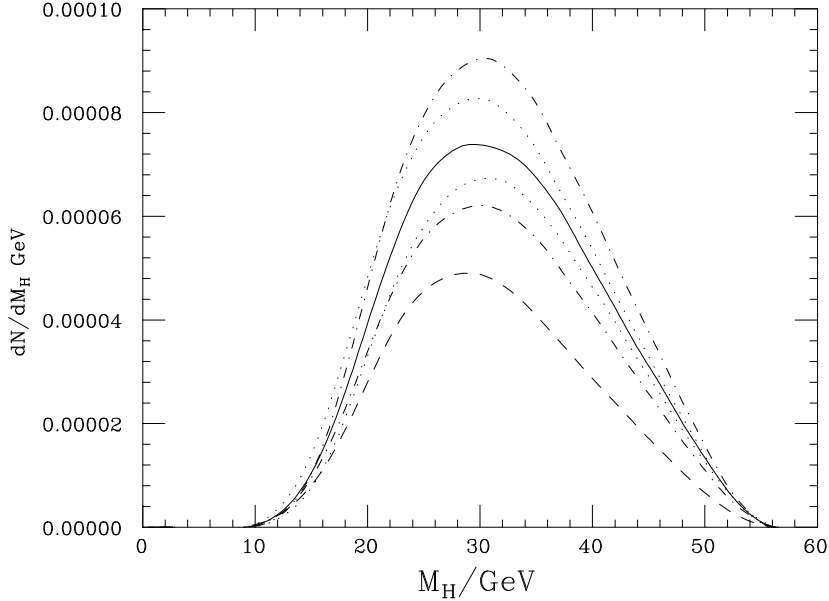}
\end{center}
\caption[bquark]{The multiplicity of bottom quark pairs as a function of
  the heavy jet mass, normalised to the number of $\Z^0$ decays.  The
  dashed curve shows the fixed order result, whereas the solid curve
  includes the resummation of the large logarithms. Also shown are the
  results of varying the quark mass by $5\%$ (dotted) and
  $\Lambda_{\mathrm{QCD}}$ by a factor of two (i.e.\  $\as$ by $10\%$)
  (dash-dotted).}
\label{fig:bquark}
\end{figure}

Of course, these shapes are dependent on the values chosen for the
parameters, $\Lambda_{\mathrm{QCD}}$ and $m_{\mathrm{Q}}$. The effect of
varying these parameters is seen in Fig.~\ref{fig:bquark}.
Also shown is the contribution from the fixed order term alone,
demonstrating the importance of resumming large logarithms.

\subsection*{Event Generators}

Monte Carlo event generators predict quite a wide range for the rate of
secondary heavy quark production.  While JETSET~[\ref{JETSET}] and
HERWIG~[\ref{HERWIG}], which are both based on the parton shower
formalism, are in quite good agreement, ARIADNE~[\ref{ARIADNE}], based
on the dipole cascade model, lies well above them.  At present the data
for the total rate lies between the two, in agreement with both,
although in somewhat better agreement with ARIADNE\@.

In Ref.~[\ref{Mike}] it was argued that the discrepancy between the
models is actually due to a specific problem with ARIADNE~-- the fact
that it allows very low transverse momentum gluons to be very massive.
If that is the case, it should show up in the distributions calculated
in the previous section.  Following the suggestion of Ref.~[\ref{Mike}],
later versions of ARIADNE have had an option\footnote{This option is
  switched on by setting \verb+MSTA(28)=1+.} to veto gluon splitting
with $m_g>k_{\perp g}$, which should fix this problem.

The parton shower and dipole cascade models are only formally accurate
to leading logarithm and do not include the exact matrix elements for
$\mathrm{q\bar{q}Q\bar{Q}}$ production.  Therefore our calculation is
more accurate than them and can be used to check them.

\begin{figure}[thbp]
\begin{center}
~\epsfig{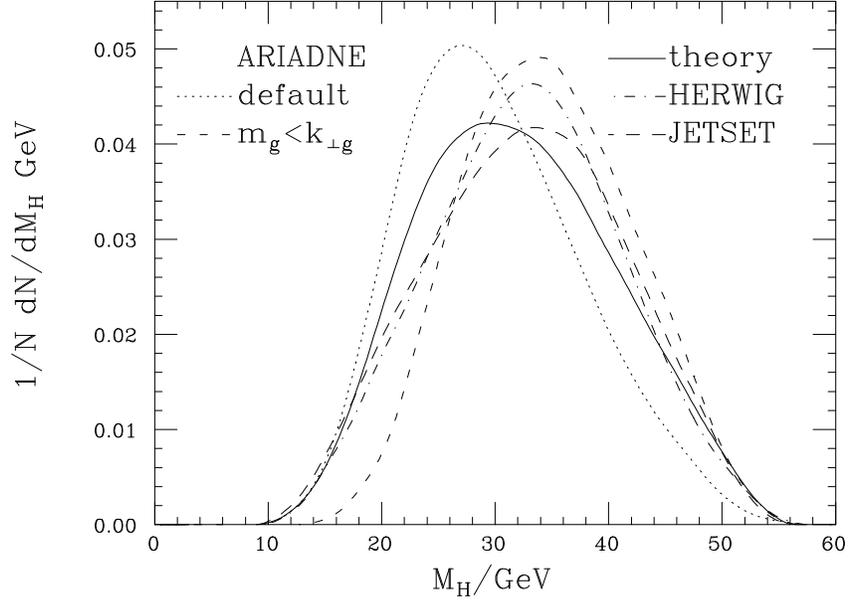}
\end{center}
\caption[MC91]{The multiplicity of bottom quark pairs as a function of
  the heavy jet mass in $\Z^0$ decays from our calculation (solid) and
  from various Monte Carlo models.}
\label{fig:MC91}
\end{figure}

In Fig.~\ref{fig:MC91} we compare our results with the predictions from
the event generators for $\surd s=m_\Z$.  With the exception of the gluon
splitting option in ARIADNE, we keep all model parameters at their
default values.
We see that HERWIG and JETSET give similar predictions for the
distribution as well as for the rate and that the unmodified ARIADNE
peaks at somewhat lower heavy jet mass than them.  Adding the new
modification, ARIADNE's distribution is more like the other models', but
still somewhat different, particularly at low jet masses.  Our results
lie between ARIADNE and the other models.

\begin{figure}[thbp]
\begin{center}
~\epsfig{file=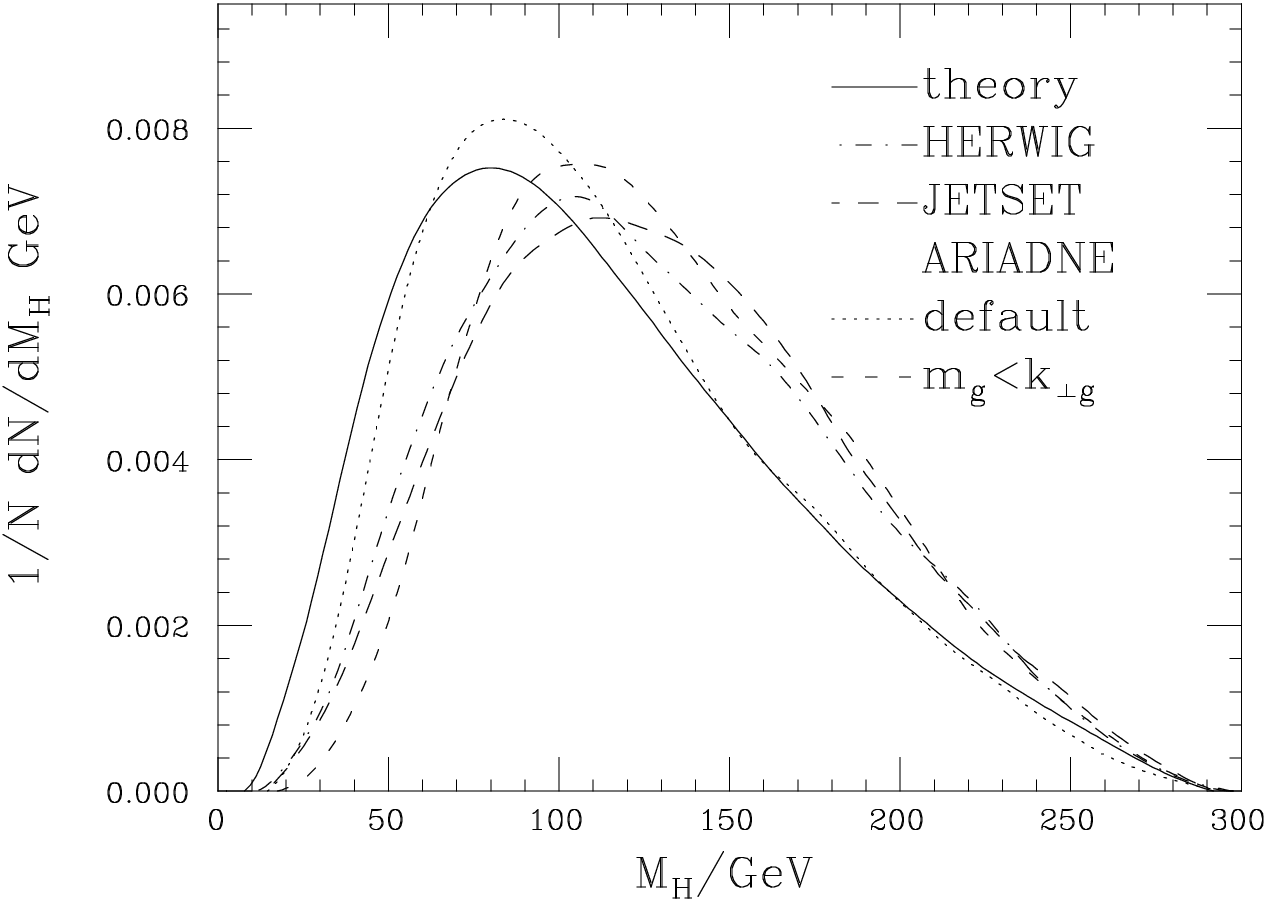,width=8cm,angle=90}
\end{center}
\vspace{-5mm}\caption[MC91]{As Fig.~\ref{fig:MC91} but at $\surd s=500$~GeV.}
\label{fig:MC500}
\end{figure}

Increasing the centre-of-mass energy, the relative importance of the
fixed order term is reduced and one gets a cleaner probe of the parton
evolution.  In Fig.~\ref{fig:MC500} we show the comparison again at
higher energy.
The modified version of ARIADNE is in even better agreement with the
other two models, while the unmodified version is in good agreement with
our calculation.  We therefore see no evidence to support the claim of
Ref.~[\ref{Mike}] that there is a problem with ARIADNE.

\subsection*{Summary}

We have calculated the multiplicity of heavy quarks from gluon splitting
in $\mathrm{e^+e^-}$ annihilation, as a function of the heavy jet mass.
Our result is exact to leading order in $\as$, and sums leading and
next-to-leading logarithms to all orders in $\as$.  
We find that the fractions of $\Z^0$ decays that
contain a secondary charm or bottom quark pair respectively, are
\be
f_\C=2.007 \%,\quad
f_\B=0.175 \%.
\ee
The shape of our
result is similar to that predicted by Monte Carlo event generators at
the $\Z^0$, lying between the different models, but in better agreement
with ARIADNE at higher energy.

\subsection*{References}

\begin{enumerate}
\item\label{Mike}
  M.H. Seymour, \np{436}{163}{95}
\item\label{Kuhn}
  A.H. Hoang, M. Je\.zabek, J.H. K\"uhn and T. Teuber,
  \pl{338}{330}{94}
\item\label{early}
  A.H. Mueller and P. Nason, \np{157}{226}{85}; \np{266}{265}{86} \\
  M.L. Mangano and P. Nason, \pl{285}{160}{92} \\
  M.H. Seymour, \zp{63}{99}{94}
\item\label{OPAL1}
  OPAL Collaboration, R. Akers {\em et al.}, \zp{67}{27}{95}
\item\label{OPAL2}
  OPAL Collaboration, R. Akers {\em et al.}, \pl{353}{595}{95}
\item\label{DELPHI}
  DELPHI Collaboration, P. Abreu {\em et al.}, \pl{405}{202}{97}
\item\label{ALEPH}
  G. Hansper for the ALEPH Collaboration, Contribution to the XXVIII
  International Conference on High Energy Physics ICHEP Warsaw, Poland,
  24--31 July 1996, \mbox{Pa~05-065}
\item\label{JETSET}
  T. Sj\"ostrand, \cpc{39}{347}{84}; \\ M. Bengtsson and T. Sj\"ostrand,
  \cpc{43}{367}{87}
\item\label{HERWIG}
  G. Marchesini, B.R. Webber, G. Abbiendi, I.G. Knowles, M.H. Seymour
  and L. Stanco, \cpc{67}{465}{92}
\item\label{ARIADNE}
  L. L\"onnblad, \cpc{71}{15}{92}
\item\label{Kniehl}
  B.A. Kniehl and J.H. K\"uhn, \np{329}{547}{90}
\item\label{CTTW}
  S. Catani, L. Trentadue, G. Turnock and B.R. Webber,
  \np{407}{3}{93}
\item\label{us}
  D.J. Miller and M.H. Seymour, `The Jet Multiplicity as a Function of
  Thrust', Rutherford Lab preprint RAL-TR-98-041, hep-ph/9805413
\item\label{CDFW}
  S. Catani, Yu.L. Dokshitzer, F. Fiorani and B.R. Webber,
  \np{377}{445}{92}
\end{enumerate}

\end{document}